\documentclass[acmsmall,10pt]{acmart}
\setcopyright{none} 
\def\BibTeX{{\rm B\kern-.05em{\sc i\kern-.025em b}\kern-.08emT\kern-.1667em\lower.7ex\hbox{E}\kern-.125emX}}
\usepackage{url}

\newcommand{\tool}{\textsc{Digger}\xspace}

\usepackage{nicefrac}
\usepackage{siunitx}
\usepackage{array,framed}
\usepackage{pdfpages}
\usepackage{booktabs}
\usepackage{
  color,
  float,
  epsfig,
  wrapfig,
  graphics,
  graphicx,
  subcaption
}
\expandafter\def\csname ver@subfig.sty\endcsname{}
\usepackage{subfig}
\usepackage{setspace}
\usepackage{amsfonts}
\usepackage{latexsym,fancyhdr}
\usepackage{enumerate}
\usepackage{algorithm2e}
\usepackage{algpseudocode}
\usepackage{graphics}
\usepackage{xparse} 
\usepackage{xspace}
\usepackage{multirow}
\usepackage{csvsimple}
\usepackage{balance}
\usepackage{bbding}
\usepackage{flushend}
\usepackage{pifont}
\usepackage{threeparttable}
\usepackage{longtable}
\usepackage{wrapfig}
\usepackage{shortcuts}
\usepackage{graphicx}
\usepackage{epstopdf}
\usepackage{url}
\usepackage{subcaption}
\usepackage{
  tikz,
  pgfplots,
  pgfplotstable
}

\usetikzlibrary{
  shapes.geometric,
  arrows,
  external,
  pgfplots.groupplots,
  matrix
}

\pgfplotsset{compat=1.9}

\usepackage{mathtools,}

\DeclareMathAlphabet{\mathcal}{OMS}{cmsy}{m}{n}

\DeclareGraphicsExtensions{%
    .png,.PNG,%
    .pdf,.PDF,%
    .jpg,.mps,.jpeg,.jbig2,.jb2,.JPG,.JPEG,.JBIG2,.JB2}

\begin{document}

\title{\tool{}: Detecting Copyright Content Mis-usage in Large Language Model Training}

\author{Haodong Li}
\affiliation{
	\institution{Beijing University of Posts and Telecommunications}
	\country{China}
	}

\author{Gelei Deng}
\affiliation{
	\institution{Nanyang Technological University}
	\country{Singapore}
 }

\author{Yi Liu}
\affiliation{
	\institution{Nanyang Technological University}
	\country{SingaporeSingapore}
 }

\author{Kailong Wang}
\affiliation{
	\institution{Huazhong University of Science and Technology}
	\country{China}
 }

\author{Yuekang Li}
\affiliation{
	\institution{The University of New South Wales}
	\country{Australia}
 }

\author{Tianwei Zhang}
\affiliation{
	\institution{Nanyang Technological University}
	\country{Singapore}
 }

\author{Yang Liu}
\affiliation{
	\institution{Nanyang Technological University}
	\country{Singapore}
 }

\author{Guoai Xu}
\affiliation{
	\institution{Harbin Institute of Technology, Shenzhen}
	\country{China}
 }

\author{Guosheng Xu}
\affiliation{
	\institution{Beijing University of Posts and Telecommunications}
	\country{China}
 }

\author{Haoyu Wang}
\affiliation{
	\institution{Huazhong University of Science and Technology}
	\country{China}
 }

\begin{abstract}
\textbf{Abstract}:Pre-training, which utilizes extensive and varied datasets, is a critical factor in the success of Large Language Models (LLMs) across numerous applications. However, the detailed makeup of these datasets is often not disclosed, leading to concerns about data security and potential misuse. This is particularly relevant when copyrighted material, still under legal protection, is used inappropriately, either intentionally or unintentionally, infringing on the rights of the authors.

In this paper, we introduce a detailed framework designed to detect and assess the presence of content from potentially copyrighted books within the training datasets of LLMs. This framework also provides a confidence estimation for the likelihood of each content sample's inclusion. To validate our approach, we conduct a series of simulated experiments, the results of which affirm the framework's effectiveness in identifying and addressing instances of content misuse in LLM training processes. Furthermore, we investigate the presence of recognizable quotes from famous literary works within these datasets. The outcomes of our study have significant implications for ensuring the ethical use of copyrighted materials in the development of LLMs, highlighting the need for more transparent and responsible data management practices in this field.
\end{abstract}

\maketitle
\section{Introduction}\label{sec:intro}
Large language models (LLMs), such GPT-2, LLaMA and GPT-3, have already achieved impressive performance in tasks such as Text generation, Text classification, and Translation. This has led to more and more people looking at and experiencing what LLM has to offer. According to a report by The Guardian, ChatGPT reached a user base of 100 million, two months after its launch\cite{chatgpt-user}.

One of the driving factors behind the prowess of these models is their extensive training on vast datasets. Such datasets, sourced from web pages, social media, and various online databases, often encompass a plethora of content, including copyright material from books. The exact constitution of LLMs' training datasets remains elusive due to either their proprietary nature or access restrictions. Notably, there have been instances in the community where LLMs have inadvertently divulged copyright-protected content~\cite{chang2023speak}, underscoring the need for an effective mechanism to detect copyright infringements during the LLM training phase.

Our methodology stems from a fundamental premise: The LLM inference result should more robustly reflect its training material after the fine-tuning process. We can potentially compare the LLM inference result with the target copyright material to determine if it is within the training data. 
With this principle, we hypothesize that a comparative analysis between the LLM's inference output and a given copyrighted content could potentially indicate the presence of such content within the training dataset. 
To validate this point, we conduct an exploratory study of sample loss dynamics using GPT-2~\cite{radford2019language}, one of the most representative open-source large language models. Our empirical observations lead to several pertinent conclusions. 
First, as training evolves, the discrepancy between inference results and the original training data diminishes, though this reduction plateaus over time. This reveals that the language loss difference base on sample loss~\cite{shokri2017membership,yeom2018privacy,song2020information,mireshghallah2022quantifying} between two LLMs on certain materials is a good indicator to identify if the material is used for fine-tuning. 
Second, model size plays a significant role: larger architectures exhibit a more rapid convergence during the initial training epochs, suggesting a faster adaptation to the target material. 
Finally, the inherent nature of the training material influences the loss. This diversity underscores the challenge of establishing a universally applicable loss criterion to reliably determine the presence or absence of specific content within disparate training corpora.

Our observations indicate that solely observing an LLM's inference outcomes on copyrighted material does not conclusively indicate its prior training on that material. Distinct target materials can yield varying inference results, even when trained on an identical LLM. Instead of this direct observation, we suggest a nuanced strategy involving the generation of a baseline dataset and simultaneous fine-tuning of the LLM using this baseline and the target dataset.

Building upon this idea, we propose \tool{}, a framework designed to discern material usage during an LLM's training. Central to our approach is the ``loss gap''—a shift in loss distribution observed during copyright detection fine-tuning. 
Initially, we generate a baseline dataset of unencountered content and establish a \textit{baseline LLM} by fine-tuning the given LLM for copyright detection. 
Concurrently, we fine-tune the target LLM with the target dataset without knowing its details. 
This \textit{baseline LLM} is then fine-tuned again with a mix of baseline dataset just learned and new dataset that is not learned, to produce a \textit{reference LLM}. By analyzing the sample loss difference between this \textit{reference LLM} and the \textit{baseline LLM}, we differentiate between familiar and new content. 
In the final step, the \textit{reference LLM} undergoes another round of fine-tuning, integrating the target dataset and new content, yielding a \textit{reference-tuned LLM}. The resulting loss gap provides insights into the target dataset's influence on inferences. 
Using this identified loss gap, we calibrate the original LLM's loss distribution, which has been fine-tuned over the target dataset. This calibrated distribution assists us in deriving a confidence score, a measure that estimates the likelihood of the target content being a part of the training materials in the initial LLM.

We rigorously test \tool{} in both controlled environments with established labels and real-world LLM scenarios. Results from our experiments on the GPT2-XL demonstrate that \tool{} achieves an accuracy of 84.750\%  and a recall of 92.428\% in the prepared LLM settings. Moreover, when we deploy \tool{} to assess if GPT2-XL and LLaMA-7b has been trained on notable quotes present in open-source training materials, it produces affirmative results, underscoring the robustness of our solution in real-world environments. To facilitate reproduction and future research, we have made our tool open-source as an anonymous project\footnote{\url{https://anonymous.4open.science/r/DIGGER-86CE/}}.

In summary, we make the following contributions in this paper:
(1) We introduce and explore a novel question surrounding the potential copyright infringements arising during the LLM training process.
(2) Building on an empirical study of GPT-2, we put forth and substantiate the theory that post fine-tuning inference behavior can shed light on the LLM's foundational training material. From this, we propose our \tool{} framework, utilizing the `loss gap' principle to pinpoint content exposure during LLM training.
(3) We demonstrate the robustness and efficacy of \tool{} through rigorous testing in both controlled environments and real-world settings, affirming its value in identifying particular training constituents within state-of-the-art LLMs.

\section{Background}\label{sec:background}
\textbf{Complications of AI Models Trained on Copyrighted Content.}
The training of artificial intelligence (AI) models, particularly LLMs, on copyrighted content has emerged as a complex issue straddling legal, ethical, and technological domains. The architecture of these AI models, especially those based on deep learning, requires the consumption of extensive datasets, often comprising copyrighted material like texts, images, or audio. This necessity raises intricate questions about traditional copyright law, as the self-generative nature of AI models makes it difficult to delineate how much of the model's outputs can be considered derivative of the original copyrighted works. High-profile legal cases such as Getty Images accessed via Stability AI~\cite{stability-ai} have brought these complexities to the forefront, demonstrating the legal risks of using copyrighted content in AI training without explicit consent. Generative AI models, capable of creating new content, add another layer of complexity. These models learn from the patterns and structures in their training data and can generate new data that may closely resemble copyrighted works. This poses significant legal and ethical challenges, including potential copyright infringement and the ethical implications of using copyrighted material without consent. The risks are particularly high when the generated content bears watermarks or other insignia, which could be interpreted as an admission of infringement.

\textbf{Limitations of Exisiting Mitigations.}  Various mitigation strategies have been proposed, ranging from legal solutions like securing explicit content use consent through contracts or licenses, to technological approaches like creating original datasets based on real-world measurements.
Despite these mitigation strategies, challenges persist. Legal solutions often involve complex negotiations and may not fully address the ethical dimensions, such as potential cultural appropriation or commodification of creative works. Technological solutions, while bypassing the need for copyrighted content, require significant investments and may still fall short in capturing the richness of human-created content. In this work, we aim to explore the copyrighted content internalized within Large Language Models, seeking to understand how these models transform and potentially infringe upon copyrighted material.

\section{Characteristic Study}\label{sec:prestudy}

We explore an intuitive approach for detecting possible copyright infringements within LLMs. Our premise is rooted in the assumption that if an LLM reproduces exact content from its training data when prompted with related content, then there is a significant likelihood that the original copyrighted material was part of its training set. To support this approach, we initially undertake an characteristic study, aiming to discern the behavioral differences of LLMs when exposed to materials they have encountered during training versus those they have not. 

Our strategy starts with examining the sample loss~\cite{yeom2018privacy,ye2022enhanced} of LLMs trained on different materials. Language modeling(LM) loss~\cite{reed1999neural,lecun2015deep,aroca2020losses} is primarily used to measure the probability that the model will generate a sequence of text, In the case of the loss function for GPT-2, the cross-entropy loss function is employed~\cite{radford2019language}, and it is represented by the following formula:

\begin{equation}
\begin{aligned}
 -\sum_{i=1}^{n} y_i \log(p_i)
\end{aligned}
\end{equation}
In this formula, $y_{i}$ represents the $i$-th element of the actual labels, and $p_{i}$ represents the $i$-th element of the model's predicted probabilities.

For the GPT model~\cite{radford2018improving}, the loss function can be represented as:
\begin{equation}
L(\Theta) = -\sum_{i} \log P(u_i | u_1, u_2, \ldots, u_{i-1}; \Theta)
\end{equation}

Here, $\Theta$ represents the model's parameters, $u_{i}$ denotes the $i$-th token in the input sequence, and $P(u_i | u_1, u_2, \ldots, u_{i-1}; \Theta)$ represents the conditional probability of predicting the next token $u_{i}$ given the previous tokens. The objective of the loss function is to minimize the negative log-likelihood.

For instance, LLaMA employ not only cross-entropy as the loss function but also make use of mean squared error (MSE)~\cite{touvron2023llama}.

Conventionally, the LM predicts the upcoming token given a specific context. Here, the loss function assesses the gap between the real token and the model's prediction. The overarching objective during LM pre-training process is to minimize this discrepancy for training dataset instances. This translates to generally lower sample losses for training dataset instances compared to non-training instances. 
By leveraging sample loss, we can discern shifts in LLM inference when it undergoes training on specific materials. With this in mind, our study centers on the following research questions:

\begin{itemize}
\item\textbf{RQ1:} How does the fine-tuning process influence the sample loss of LLM with respect to the target material?
\item\textbf{RQ2:} Can sample loss be used to identify whether a material has been learned by a material before?
\end{itemize}

In the following of this section, we first delineate our experimental design to address the research questions. Subsequently, we showcase the experimental results and our key findings.


\subsection{RQ1: Impact of Fine-tuning}
\label{sec:pre_RQ1}
To address \textbf{RQ1}, our key strategy is to finetune an open source LLM with materials that has not been learned, and observe the change in sample loss before and after the fine-tuning process. We then vary the size of the model and the learning times and compare the experimental results.

\noindent\textbf{Model Selection}.
For our study, we utilize publicly accessible, pre-trained checkpoints of GPT-2~\cite{radford2019language} sourced from the Huggingface Model Hub. A key benefit of using this model is its availability in multiple versions and its lightweight nature. We focus our investigations on text generation models, specifically those LMs pre-trained on the WebText~\cite{radford2019language} dataset, which encompasses 40GB of English content sourced from the web. Our experimental models include GPT-2 with its four size variants: Small (124m parameters), Medium (355m), Large (774m), and XL (1557m).

\noindent\textbf{Dataset Construction Selection}.
Given the uncertainty regarding the specific books included in the GPT-2 training set, we opted for a definitive approach by selecting content that could not have been part of it. To this end, we pick 5 books(Detailed information can be found in Appendix ~\ref{appendix:pre_study_book_list}.) published in 2023 for our preliminary training set. From these novels, we extract 100 unique 500-word passages. Three distinct training sets emerge, distinguished by sample repetition counts of 1, 2, and 3. For testing, the initial 100 tokens from each sample serve as our data. 

\begin{figure}[]
\centering
\includegraphics[width=0.98\textwidth]{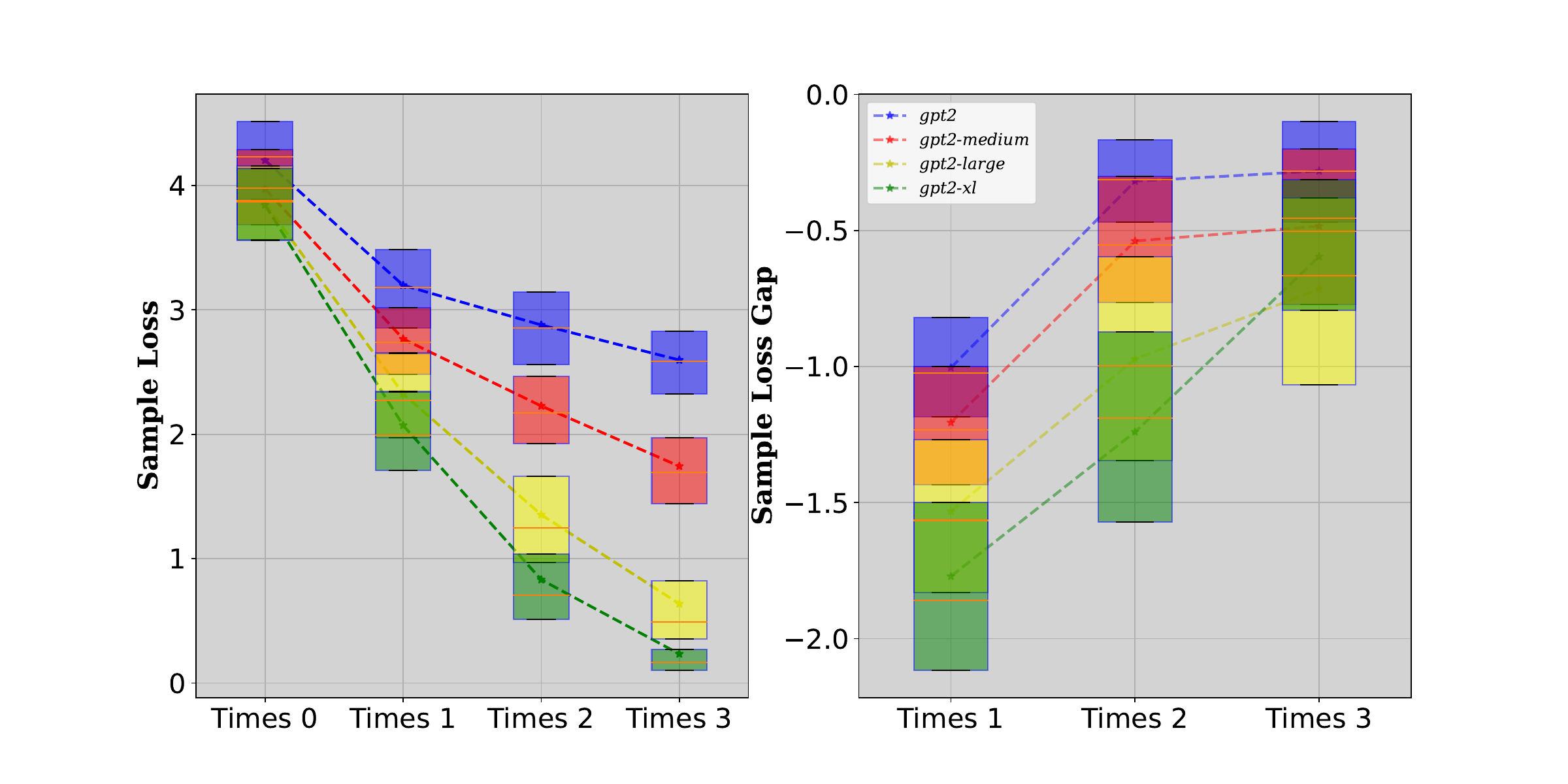}
\caption{Different versions of GPT-2 have different numbers of samples learning iterations. The changes in sample loss on the GPT-2 and the magnitude of the loss variation are as follows.}
\label{fig:pre_study_loss}
\end{figure}

\noindent\textbf{Experimental Results and Analysis.} We present the experimental results in Figure~\ref{fig:pre_study_loss}. As demonstrated in the left subfigure of Figure~\ref{fig:pre_study_loss}, it is evident that the loss an LLM incurs for a given sample diminishes with the increasing frequency of that sample in the LLM's training set. The base LLM, which has not been exposed to these samples, exhibits the highest loss. Conversely, when the LLM has encountered these samples three times in its training, it registers the lowest loss. This observation underscores the notion that the more frequently an LLM is trained on a specific sample, the more optimized its performance becomes, as indicated by the reduced loss. 

The right subfigure of Figure~\ref{fig:pre_study_loss} illustrates the rate of change in sample loss. It is observed that the first-time exposure to the training material yields the steepest loss change, while subsequent trainings witness a tapering loss gradient. This trend implies that we can rely on the magnitude of the change in loss to distinguish whether LLMs has learned some samples or not. 

Combining the expeirmental results on different models, we also notice that larger models achieves quicker convergence during the training process: Compared to GPT-2, the sample loss drop of GPT-2-XL is faster during the training process, with quicker convergence rate. 

\begin{framed}
\noindent \textbf{Key Findings: When the LLM undergoes repeated learning on specific samples, there's a noticeable trend of diminishing sample loss. This decline, however, is not linear - as the number of learning repetitions increases, the rate of loss reduction starts to taper off. Among the models analyzed, GPT2-XL stands out, showcasing the most rapid convergence during its training phase.}
\end{framed}

\subsection{RQ2: Evaluation Metrics Investigation}

From our observations in \textbf{RQ1}, the reduction in sample loss emerges as a promising signal to discern if an LLM has been trained over specific material. However, this decrement in loss resulting from LLM fine-tuning is influenced by both the model's architecture and the training content. Hence, prescribing a singular loss threshold to ascertain if a material is part of an LLM's training set is not feasible. In \textbf{RQ2}, we propose to use the strategy of creating ``reference LLM'' to fairly detect if a material has been learned by the target LLM before.  

\begin{figure}[]
\centering
\includegraphics[width=0.98\textwidth]{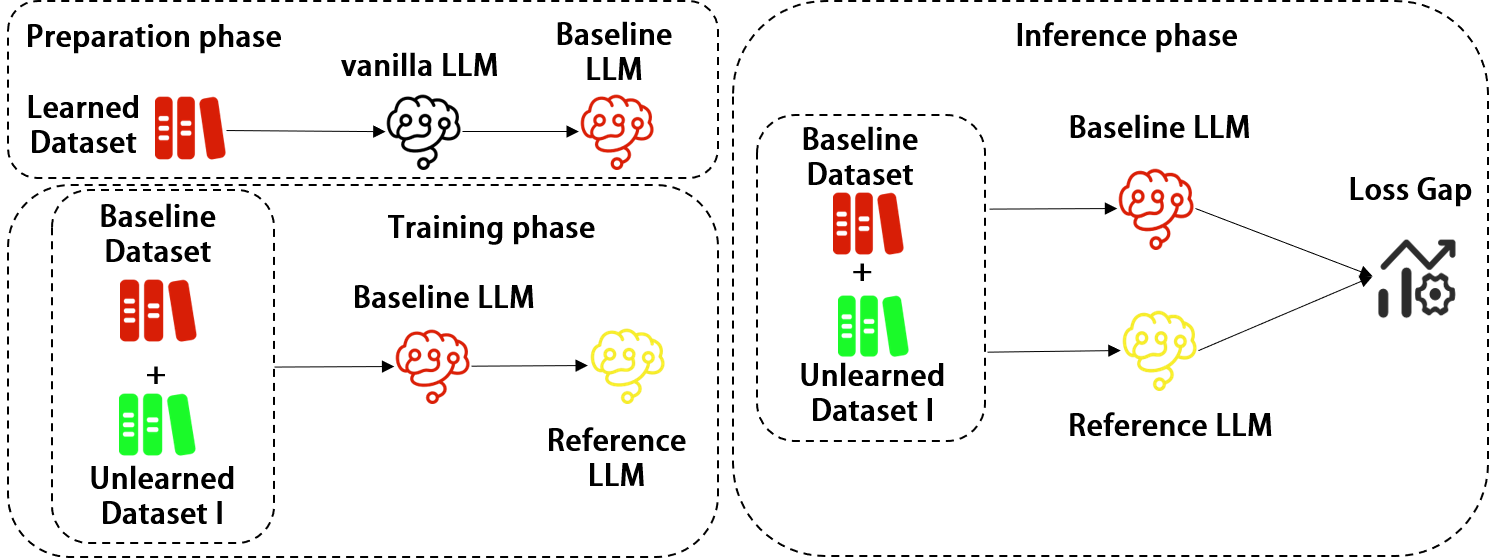}
\caption{ Overview of Our preliminary study consists of three phases, namely, 1)Preparation phase, 2) Training phase and 3) Inference phase. LLMs marked with different colors indicate that it was finetuned with different datasets.}
\label{fig:pre_study_framework}
\end{figure}

\noindent\textbf{Establishing a Ground Truth.} As illustrated in Figure~\ref{fig:pre_study_framework}, our primary objective is to differentiate between datasets that the LLM has been trained on and those it has not encountered. Directly discerning which materials an LLM has previously been exposed to is a challenging endeavor. To navigate this challenge, we initiate a \textit{preparation phase} to create a \textit{target LLM}. We begin by selecting novels that were published after the initial development of the LLM, ensuring they have not been used in its training. These novels are divided equitably into two distinct categories: the \textbf{learned set} and the \textbf{unlearned set}. By fine-tuning the original LLM using the \textbf{learned set}, we generate a target LLM that is definitively trained on specific materials. Our subsequent goal is to differentiate between the learned and unlearned sets by evaluating sample loss.

\noindent\textbf{Experimental Framework.} With our ground truth established, our experimental procedure is depicted in the Training and Inference phases of Figure~\ref{fig:pre_study_framework}. During the training phase, we merge the learned and unlearned datasets, using this combined set to train the \textit{target LLM}, which results in the formation of the \textit{reference LLM}. In this context, the learned set is exposed multiple times to the reference LLM during training, whereas the unlearned set is trained only once. In the inference phase, we compute the sample loss of the integrated dataset using both the target LLM and the reference LLM, producing two separate loss distributions. Based on insights from \textbf{RQ1}, we anticipate a noticeable difference between these two distributions.

\noindent\textbf{Evaluation Metrics.}
Our primary evaluation metric is the Area Under the ROC Curve (AUC). The ROC curve, as described in the cited work~\cite{metz1978basic,fawcett2006introduction}, illustrates the relationship between power (true positive rate) and error (false positive rate) at various thresholds $t$ chosen for the membership prediction process. It effectively encapsulates the balance between power and error. AUC is crucial in evaluating binary classification tasks as it not only provides a measure of the task's efficacy but also assists in identifying optimal thresholds.

\noindent\textbf{Experiment Settings.} We adhere to the experimental framework outlined in \textbf{RQ1}, employing the four GPT-2 variations as the training models. To introduce variability in the preparation phase, the base LLM is exposed to the dataset one to four times, consecutively, resulting in four distinct target LLMs for comparison. During the inference phase, we utilize samples of varying lengths, from 50 to 100 tokens, with intervals of 10 tokens, to analyze the influence of token length on loss.

\noindent\textbf{Experimental Results.} Table~\ref{tableRQ1} delineates our findings. Specifically, it displays the AUC values for various models across different settings. For GPT-2, when samples appear once in the training set, the lowest AUC (0.67318) is recorded with test samples comprising 50 tokens. This value peaks at 0.78235 for test samples with 100 tokens. A similar trend is observed with other models. For instance, with XL and a single sample repetition, the minimum AUC is 0.89705 at 50 tokens, while the maximum is 0.96670 at 100 tokens.

For the Medium configuration with test samples fixed at 60 tokens and training samples repeated once, the AUC is at a minimum of 0.79122. With three repetitions, this value increases to 0.97928. As the repetition frequency of samples escalates, there is a general uptrend in the AUC. For example, in the XL configuration with 100-token test samples and one repetition in training, the AUC is 0.96670. However, with three repetitions, this value nearly reaches perfection at 0.99995.

When test samples are comprised of 100 tokens and appear once in the training set, the corresponding AUC values for GPT-2 Medium, Large, and XL are 0.78235, 0.87429, 0.95222, and 0.99995, respectively. This suggests a correlation between the LLM's parameter size and the AUC: as the former increases, the latter tends to rise, assuming constant token lengths in test samples and training sample repetitions.

\begin{table}[H]
    \centering
    \caption{Classification effects (AUC) for two samples with different GPT-2 versions, different token lengths, and different number of repetitions of the samples. The best results are shown in bold.
    }
     {
  
    \begin{tabular}{cc|cccccc}
    \hline
     \multicolumn{2}{c|}{\textbf{AUC}}  &\multicolumn{6}{c}{\textbf{Number of Token}}  \\
     \hline{}

    \multirow{2}{*}{\shortstack{\textbf{Version}}} &\textbf{repeat}&\multirow{2}{*}{\textbf{50}} & \multirow{2}{*}{\textbf{60}}  & \multirow{2}{*}{\textbf{70}} & \multirow{2}{*}{\textbf{80}} & \multirow{2}{*}{\textbf{90}}& \multirow{2}{*}{\textbf{100}}\\
     &\textbf{times}\\
    \hline
    \hline
   
    \multirow{3}{*}{\shortstack{GPT-2}} 
    &1  &0.67318  &0.70111 &0.72455 &0.74608 &0.76583 &0.78235\\
    &2  &0.76828  &0.80316 &0.83085 &0.85447 &0.87472 &0.89077\\
    &3  &0.84160  &0.87639 &0.90219 &0.92249 &0.93864 &0.95047\\
    
    \hline
    
    \multirow{3}{*}{\shortstack{Medium}} 
    &1  &0.75657  &0.79122 &0.81788 &0.84062 &0.85942 &0.87429\\
    &2  &0.89324  &0.92352 &0.94312 &0.95730 &0.96767 &0.97433\\
    &3  &0.96460  &0.97928 &0.98708 &0.99165 &0.99442 &0.99619\\

    \hline
   
    \multirow{3}{*}{\shortstack{Large}} 
    &1  &0.86596  &0.89626 &0.91749 &0.93277 &0.94408 &0.95222\\
    &2  &0.98733  &0.99291 &0.99532 &0.99673 &0.99748 &0.99804\\
    &3  &0.99919  &0.99952 &0.99964 &0.99969 &0.99974 &0.99975\\
    
    \hline
    
    \multirow{3}{*}{\shortstack{XL}} 
    &1  &0.89705  &0.92303 &0.93964 &0.95218 &0.96107 &0.96670 \\
    &2  &0.99718  &0.99845 &0.99893 &0.99908 &0.99928 &0.99940 \\
    &3  &0.99989  &0.99989 &0.99990 &0.99990 &0.99991 &0.99995 \\

    \hline
   
        \end{tabular}}
    \label{tableRQ1}
\end{table}

\noindent\textbf{Conclusion}
In our study, the Area Under the Curve (AUC) metric was employed to assess the efficacy of distinguishing samples that the LLM had previously encountered from those it had not. The highest AUC value recorded was 0.99995, which corresponds to the XL model where training samples were reiterated thrice, and the test samples were of a 100-token length. On the other end of the spectrum, the lowest AUC value, 0.67318, was associated with the GPT-2 model where training samples were presented only once and test samples were 50 tokens in length. These findings suggest that LLMs with a greater parameter size, when exposed to training samples more frequently, demonstrate an enhanced retention capability. Moreover, test samples with extended token lengths seem to be more effective in eliciting the stored knowledge of LLMs.

\begin{framed}
\noindent \textbf{Key Findings: LLMs with larger architectures, such as the XL model, demonstrate superior retention when trained with repeated samples. Test samples with more tokens enhance this effect, suggesting both LLM size and token length significantly impact the model's ability to recall prior knowledge. Yet, setting a universal loss threshold to gauge this remains elusive due to the nuanced interplay of LLM structure and training content.}
\end{framed}

\section{Methodology}\label{sec:methodology}
\begin{figure}[]
\centering
\label{fig:Characteristic_Study}
\begin{subfigure}[t]{0.95\textwidth}
\includegraphics[width=\textwidth]{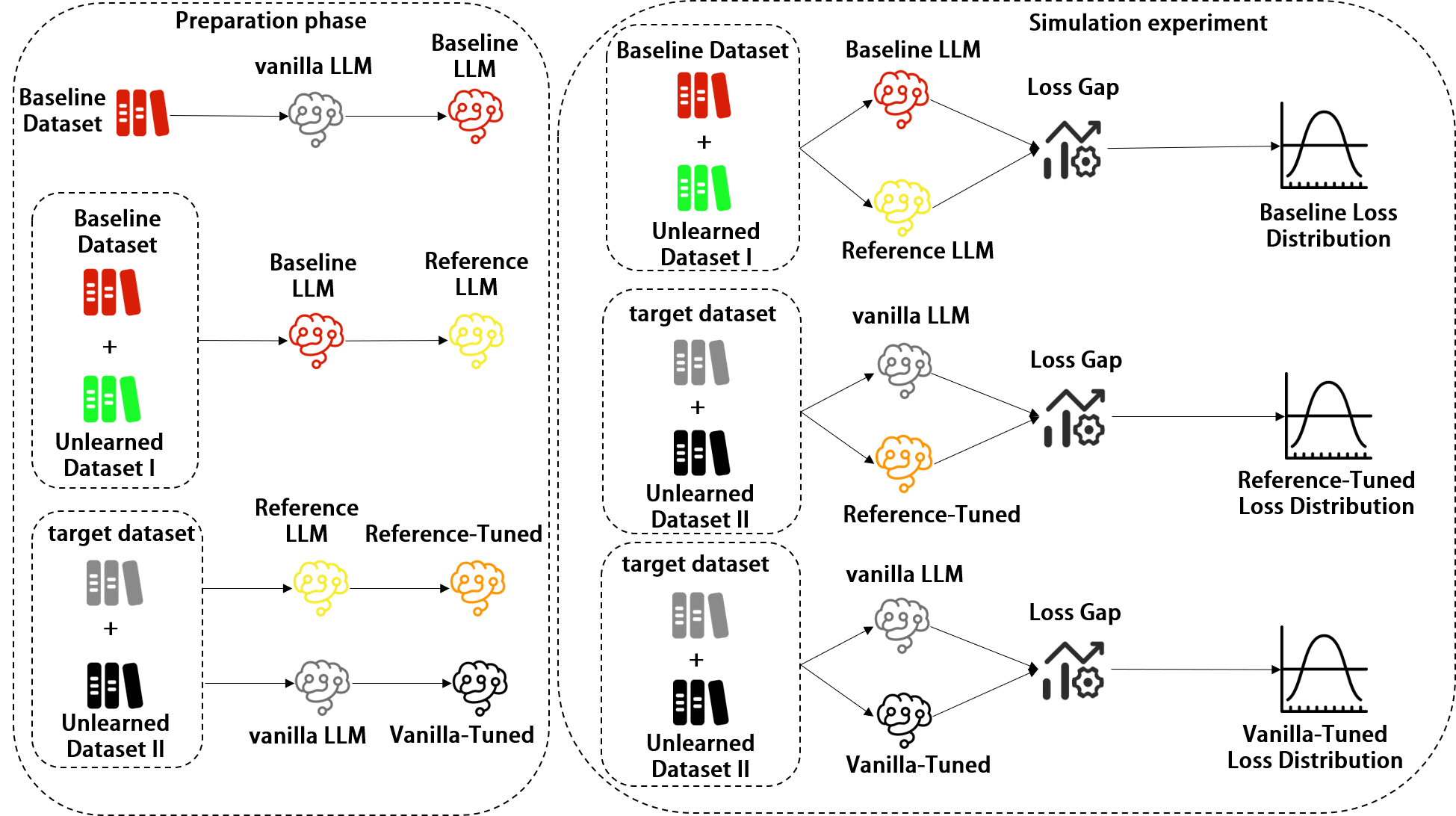}
 \label{fig:method_up}
\end{subfigure}
\vspace{-0.1in}
\begin{subfigure}[t]{0.94\textwidth}
\includegraphics[width=\textwidth]{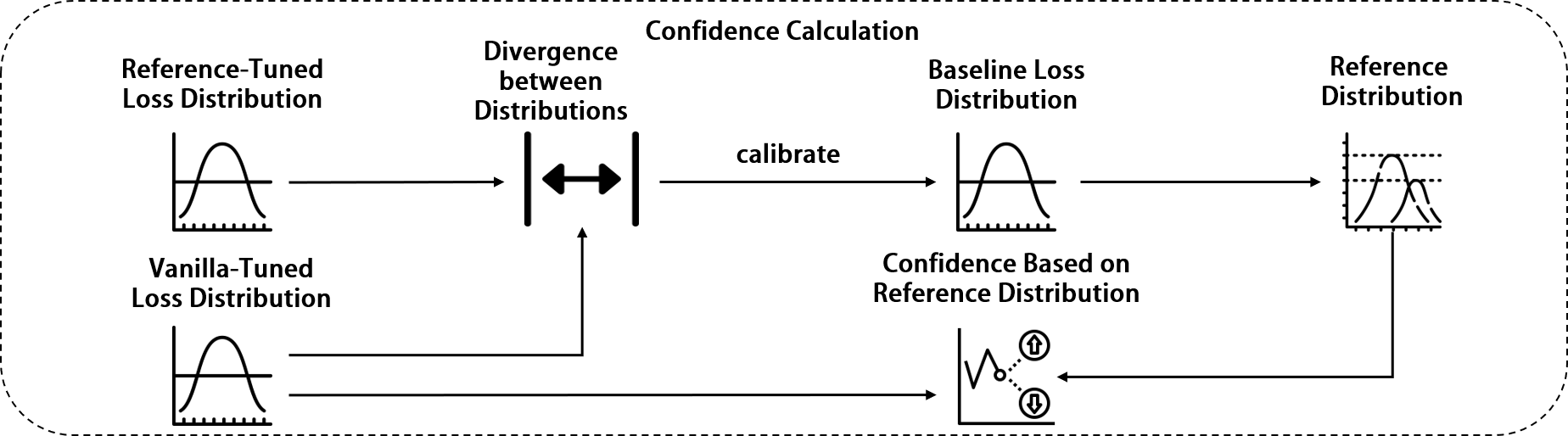}
\label{fig:method_down}
\end{subfigure}
\caption{Overview of Our Methodology consists of three phases, namely, 1) Preparation phase , 2) Simulation Experiment phase, and 3) Confidence Calculation phase. LLMs marked with different colors indicate that it was finetuned with different datasets.}
\label{fig:method}
\end{figure}

Enlightened by the findings of preliminary study, we propose \tool, a general framework to identify if the given target material has been trained on the a given LLM. Figure\ref{fig:method} illustrates the overall workflow of the framework, which contains three main phases. Given the target dataset for copyright detection and  the vanilla LLM as the target, we complete the LLM copyright detection in three phases. (1) \textbf{Preparation}, in which we create the reference LLMs as the preparation work. 
(2) \textbf{Simulation Experiment}, in which we obtain the sample loss of datasets over various trained LLMs. This step is used to understand the exact sample loss difference that will be caused by unkonwn target dataset during the fine-tuning process. 
(3) \textbf{Confidence Calculation}, in which we calculate the exact confidence score based on the sample loss differences of the target dataset over the baseline model and the reference model. 

In the following of this section, we detail the design and implementation for each phase and demonstrate the performance of our solution through experiments.


\subsection{Training phase}\label{Training_phase}
Initiating our approach based on the methodology detailed in \textbf{RQ2}, we commence by training a foundational Large Language Model (LLM) using a known, learned dataset. Subsequent to this, we cultivate the reference LLM through an additional fine-tuning process which integrates the foundational dataset with a previously unlearned dataset. These processes yield two instrumental benchmarks, the baseline and the reference LLMs, which facilitate a nuanced understanding of the loss differentials between learned and unlearned data.

In the ensuing steps, we subject the reference LLM and the vanilla LLM to fine-tuning using the target dataset, resulting in the formation of two distinct LLMs: the \textit{reference-tuned LLM} and the \textit{vanilla-tuned LLM}. It is pertinent to note that we introduce an equal measure of unlearned data to the target dataset, ensuring a harmonized juxtaposition with the foundational dataset's proportion in preceding configurations.

For dataset creation, we curate novels — assuredly not part of the target LLM's training — to assemble the foundational and unlearned datasets. Specifically, the entire novel collection is trisected to formulate three distinct datasets: foundational, unlearned I, and unlearned II. From each novel within these categories, we extract $N$ passages of a consistent length, $l$, thus acquiring the datasets requisite for the preparatory phase.

\subsection{Simulation Experiment Phase}\label{SimulationExperimentPhase}

The simulation experiment phase is paramount to the workings of \tool{}. Herein, our primary objective orbits around discerning the dynamics of loss distribution across varied LLM architectures. The aspiration is to unravel the intricate transformations in loss distribution post-fine-tuning, particularly when an LLM is exposed to an unfamiliar target dataset. Such insights would allow for juxtapositions of tangible training results against an ideal benchmark, facilitating determinations regarding the target dataset's exposure to training. The intrinsic challenge here lies in the absence of target dataset labels. Thus, contrasts are predominantly feasible with standard datasets, as articulated in \textbf{RQ2}. Yet, it's noteworthy that the discerned distributional fluctuations are tethered to our chosen foundational dataset. To mitigate this differential, we propose a reference dataset paradigm, as depicted in Figure~\ref{fig:method}.

As delineated in Figure~\ref{fig:method}, three discernible sample loss distribution categories emerge: \textit{vanilla-tuned}, \textit{reference-tuned}, and \textit{baseline}. The \textit{vanilla-tuned loss distribution}, derived from labeled foundational datasets, resonates with insights from \textbf{RQ2}, offering a robust benchmark contrasting learned and unlearned datasets. In a subsequent trajectory, the \textit{reference-tuned loss distribution} materializes when the reference LLM is fine-tuned using a blend of the target and select unlearned samples. This pivotal distribution casts light upon loss disparities introduced by the incorporation of foundational and target datasets. Contrastingly, the \textit{baseline loss distribution} is birthed when the target dataset singularly fine-tunes the vanilla LLM. As this distribution's divergence from the \textit{reference-tuned} variant encapsulates the foundational dataset's imprint on the vanilla LLM, this differential becomes a pivotal calibration tool for the \textit{vanilla-tuned loss distribution}.

\subsection{Confidence Calculation}
Transitioning from the derived loss distributions, our focus shifts towards ascertaining if the furnished target dataset exists within the specified LLM, a strategy vividly portrayed in Figure~\ref{fig:method}. Employing the \textit{Reference-tuned loss distribution} in tandem with the \textit{Vanilla-tuned loss distribution}, we compute their distributional divergence. 

In order to quantify the dissimilarity between two distributions, we employ the Wasserstein distance~\cite{tong2021diffusion,zhao2019moverscore,van2021approximating,gardner2017definiteness}. It is a mathematical method used to measure the distance between two probability distributions. It quantifies the minimum average cost required to transform one distribution into another. The computational formula for it is as follows:

\begin{equation}
W(P, Q) = \inf_{\gamma \in \Pi(P, Q)} \int_{\mathbb{R}^d \times \mathbb{R}^d} \|x - y\| \, d\gamma(x, y)
\end{equation}

Where:
\begin{itemize}
  \item $W(P, Q)$ represents the Wasserstein distance between probability distributions $P$ and $Q$.
  \item $\inf$ denotes taking the infimum over all joint distributions $\gamma$ whose marginals are $P$ and $Q$.
  \item $\Pi(P, Q)$ is the set of all joint distributions $\gamma$ with marginals $P$ and $Q$.
  \item $\|x - y\|$ represents the Euclidean distance (or an appropriate distance metric) measuring the distance from point $x$ to point $y$.
  \item $d\gamma(x, y)$ represents the differential measure of the joint distribution $\gamma$.
\end{itemize}

The Wasserstein distance intuitively measures the amount of work needed to transform one probability distribution $P$ into another distribution $Q$ with the minimum total transportation cost. It finds applications in various fields such as probability distribution matching, image processing, and natural language processing.

This resultant distance is subsequently harnessed to calibrate the \textit{vanilla-tuned} loss distribution, ultimately yielding a standard reference distribution that proves invaluable for adjudicating membership inferences concerning the target dataset. Adhering to the blueprint laid out in \textbf{RQ2}, we employ the AUC to determine a loss threshold $t$ that ensures an acceptable false-positive rate $p$. With this, we can contrast the sample losses of the target dataset against the threshold $t$ on the vanilla LLM to ascertain prior training on the LLM.

Having established our methodology, we proceed to its empirical validation. In the following section, we will present real-world experiments to demonstrate the effectiveness and applicability of our proposed loss threshold in discerning prior training on the target LLM.

\section{Evaluation}\label{sec:experiments}
We further set up experiments, aiming to assess the capability of \tool{} in determining whether samples belong to the training set of the vanilla LLM. Initially, we validate \tool's effectiveness in a controlled experimental environment before its real-world application. In particular, we aim to resolve the following two research questions.

\begin{itemize}
\item \textbf{RQ3:} How effective is \tool{} in identifying samples that belong to the training set of the vanilla LLM?
\item \textbf{RQ4:} Can \tool{} work effectively on real-world LLMs without labels?
\end{itemize}

\subsection{Experiment Setup}
Following the proposed methodology, we implement our framework using the Pytorch framework with a total number of 3,000 line of code. For the fine-tuning process involved, we use GeForce A100 with a VRAM capacity of 40GB and driven by CUDA.

\subsubsection{Building Datasets}
We collect 120 novels from the real world to form our experimental dataset. In particular, we select novels from Z-Library\cite{Z-Library}, filtered by the requirements of published after 2023. This guarantees that the tested LLM has not learned the content of these novels.
After this, we divide the collected novels into 4 different datasets:

\begin{itemize}

\item \textbf{Baseline Dataset}: Comprising 35 novels, this dataset emulates novels previously learned by the vanilla LLM. Metrics from samples within this set shape the seen sample distribution in the \textit{Vanilla-Tuned Loss Distribution}.

\item \textbf{Unlearned Dataset I}: Mirroring the \textit{Baseline Dataset} in size, this dataset represents novels not previously learned by the vanilla LLM. Its sample metrics contribute to the unseen sample distribution in the \textit{Vanilla-Tuned Loss Distribution}.

\item \textbf{Target Dataset}: Designed to test the familiarity of vanilla LLM with specific samples, this dataset has three versions during the \tool evaluation. Each version comprises 15 novels, either entirely familiar to the vanilla LLM, completely unfamiliar, or a balanced mix of both. In real-world evaluations, it includes 500 excerpts sourced from Goodreads~\cite{quote}, potentially featuring passages the vanilla LLM may have come across during training.

\item \textbf{Unlearned Dataset II}: Equivalent in size to the target dataset, this dataset contains content unfamiliar to the vanilla LLM.
\end{itemize}

We further construct as detailed in Section~\ref{Training_phase} to obtain objective measurement of our method’s effectiveness(a \textit{Baseline LLM}, a \textit{Reference LLM}, a \textit{Reference-Tuned} and a \textit{Vanilla-Tuned}) as follows:

\begin{itemize}

\item \textbf{Baseline LLM}: To obtain true labels for samples that have been learned, we finetune vanilla LLM on baseline dataset to obtain the \textit{baseline LLM}.
We designate \textit{baseline LLM} as our target LLM for the experimental phase. The contents of the \textit{baseline dataset} are reflective of its prior learning experiences.

\item \textbf{Reference LLM}: We finetune \textit{baseline LLM} to obtain the \textit{reference LLM} by using the union of the \textit{baseline dataset} and the \textit{unlearned dataset I}. 
The purpose was to establish differences in the number of learning iterations on \textit{reference LLM}. 
The \textit{baseline dataset} was learned twice by 
\textit{reference LLM} (once by \textit{baseline LLM} and once by \textit{reference LLM}), while the \textit{unlearned dataset I}  was only learned once.

\item \textbf{Reference-Tuned}:
We finetuned \textit{reference LLM} using the union of target dataset and \textit{unlearned dataset II} to obtain the \textit{reference-Tuned}.

\item \textbf{Vanilla-Tuned}:
We finetuned vanilla LLM using the union of the target dataset and \textit{unlearned dataset II} to obtain the \textit{vanilla-Tuned}.

\end{itemize}

\subsubsection{Configurations}
To explore more LLMs, other than the full series of GPT-2, we also select LLaMA-7b~\cite{touvron2023llama} as the vanilla LLM in our experiments. Following the proposed methodology section, in the \textbf{Preparation phase}, the token length of each sample was chosen to be 128. The number of training repetitions for each sample is 1, 2 and 3. In the \textbf{Inference phase}, the token lengths for test samples ranged between 50 and 100, incremented by 10.


\subsection{RQ3: The performance in the experimental environment}
\begin{figure}[]
\centering
\includegraphics[width=0.98\textwidth]{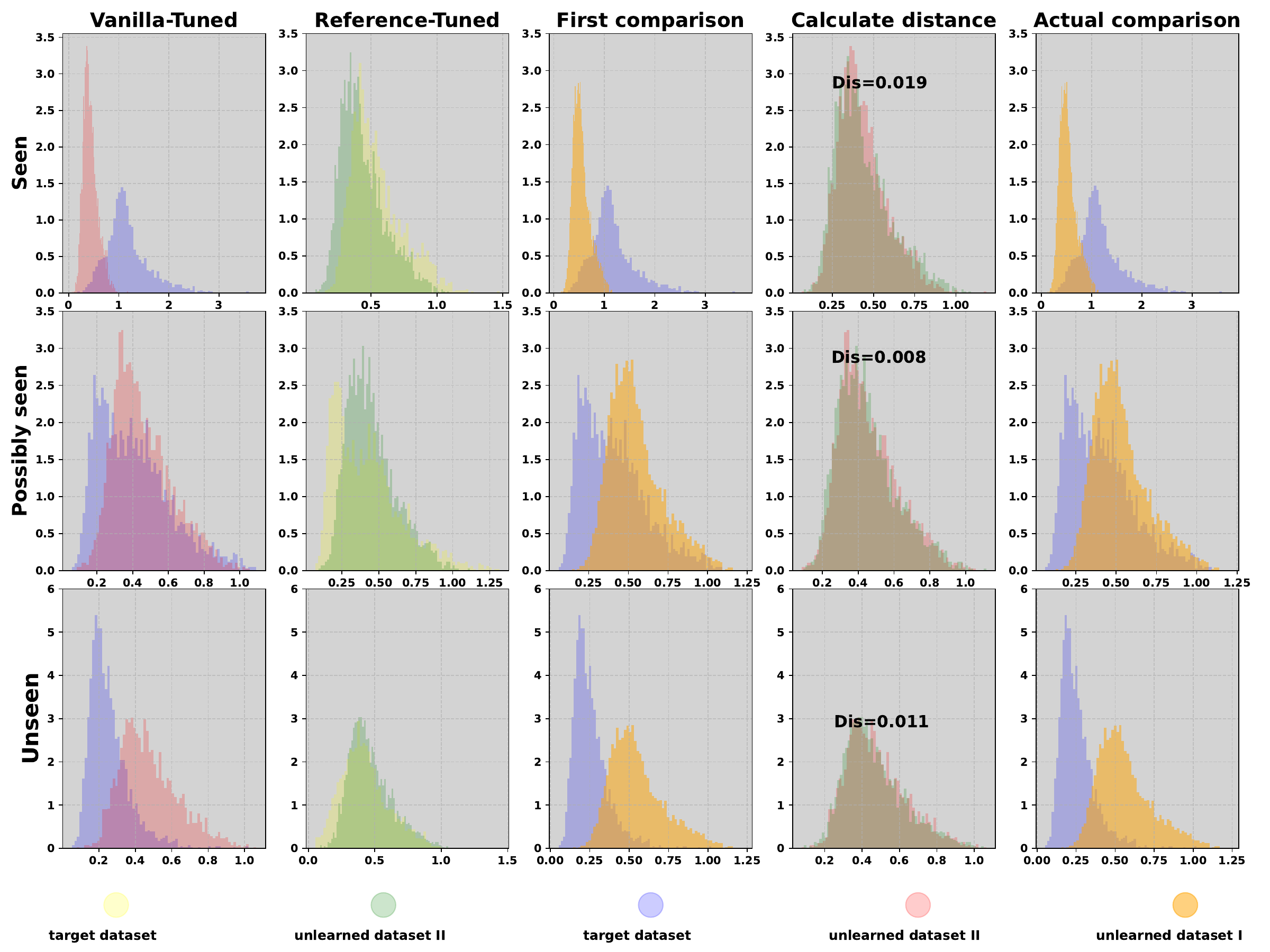}
\caption{
The figure represents distribution curve graphs for various stages. We designed three comparative experiments based on the differences in samples within the target dataset.}
\label{fig:comparsion}
\end{figure}

To rigorously assess the performance of \tool{}, we embarked on a series of controlled experiments. Specifically, we crafted three distinct target LLMs for a blind test, each representing a unique learning scenario: fully learned, entirely unlearned, and a balanced combination of both learned and unlearned content. Notably, each of these three experimental sets was conducted on a consistent baseline LLM and reference LLM. As a result, the baseline loss distribution remained invariant across all experimental groups. Yet, the dynamics of the reference-tuned loss distribution and the vanilla-tuned loss distribution fluctuated based on the nature of the target dataset. A visual representation of the distribution plots from these experiments can be observed in Figure~\ref{fig:comparsion}.

In the column marked "Vanilla-Tuned," we present three vanilla-tuned loss distribution plots, each stemming from distinct comparative experiments. When the target dataset is composed solely of seen samples (designated as the "Seen" rows), a straightforward distinction emerges between the samples in the target dataset and those in the \textit{unlearned dataset II} by merely comparing the loss gap between the vanilla LLM and \textit{vanilla-Tuned}. This distinction has been quantified with an AUC value of 0.947. However, as the proportion of unseen samples in the target dataset grows, the disparity between the distribution curves of the two sample sets diminishes. In cases where the target dataset is exclusively comprised of unseen samples (labeled "Unseen" rows), the distributions start to overlap. Occasionally, values from the target dataset even fall below the minimum values found in the unlearned dataset.

Meanwhile, the "Reference-Tuned" column showcases three reference-tuned loss distribution plots, also derived from comparative experiments. Notably, the distribution disparity between the two sample types here is subtler than what's observed in the "Vanilla-Tuned" column. Nonetheless, a discernible trend remains: as the count of unseen samples surges, the sample distribution in the target dataset progressively skews to the left, compared to the distribution in the unlearned dataset II.

To assess the sample distribution of the target dataset in the vanilla-tuned loss distribution, we leverage the sample distribution curve of the unlearned dataset I from the baseline loss distribution as a benchmark. This analysis is illustrated in the "First comparison" column of Figure~\ref{fig:comparsion}. Notably, these distributions capture the variance in sample losses for unlearned samples during their initial interaction with the LLM without prior learning.

The "Calculate distance" column contrasts the sample distributions from the unlearned dataset across both the vanilla-tuned and reference-tuned loss distributions. From our three sets of comparative experiments, we discerned distribution distances of 0.019, 0.008, and 0.011, respectively.

Subsequently, in the "Actual comparison" column, the previously calculated distribution distances are integrated into the distribution of the unlearned dataset I in the baseline loss distribution by a rightward shift. This adjusted distribution serves as a benchmark for evaluating the target dataset.

Considering the second control experiment, where the target dataset is evenly split between previously studied and novel samples, we derived ROC curves and achieved an AUC value of 0.912. Adjusting the False Positive Rate (FPR) to intervals of 5\%, 10\%, 15\%, 20\%, and 25\%, we assessed the classification results in terms of accuracy and F1 score for both sample types, as tabulated in Table~\ref{table:rq3-acc}. For instance, at an FPR of 5\%, the defined threshold is 0.460, yielding an accuracy of 69.857\%, a True Positive Rate (TPR) of 44.714\%, a False Negative Rate (FNR) of 55.286\%, and an F1 score of 59.733\%. As the FPR escalates, there's an inverse relation with the threshold. This leads to rising accuracy, F1 score, and TPR, while the FNR recedes. However, when the FPR reached 25\%, performance metrics like accuracy and F1 score began to wane.

For the first and third control experiments, the target dataset solely contains one sample type, either completely familiar or unfamiliar. In such scenarios, using ROC for evaluation isn't feasible. Instead, we classified samples based on confidence scores, distributing them across ten intervals, each spanning 0.1 units. These findings are detailed in Table~\ref{table:rq3-sample_num}. For novel samples displaying pronounced loss fluctuations, a noteworthy 83.33\% (or 2,499 samples) were located within the 0 to 0.1 confidence interval. Conversely, for samples previously encountered, demonstrating minimal loss variance, about 86.67\% (or 2,600 samples) were situated within the 0.9 to 1 confidence interval.

\begin{table}[]

    \centering
    \caption{For samples that may have been learned by the GPT2-XL, obtain a threshold controlling for the FPR of unlearned samples and the number of learned samples to be selected with this threshold
    }
     {
  
    \begin{tabular}{cc|cccc}
    \hline
     \multicolumn{2}{c|}{\textbf{GPT2-XL}}  &\multirow{2}{*}{\textbf{Acc(\%)}} &\multirow{2}{*}{\textbf{TPR(\%)}} &\multirow{2}{*}{\textbf{FNR(\%)}} &\multirow{2}{*}{\textbf{F1(\%)}}  \\
     \cline{1-2}
     
    \textbf{FPR(\%)} & \textbf{Threshold }&  \\
    \hline
  
    {\shortstack{5}} &0.460 &69.857 &44.714 &55.286 &59.733
    \\
    \hline
    {\shortstack{10}}  &0.255 &79.929 &69.857 &30.143 &77.681
    \\
    \hline
    {\shortstack{15}}  &0.176 &83.464 &82.000 &18.000 &83.219
    \\
    
    \hline
    {\shortstack{20}}&0.128 &84.750 &89.500 &10.500 &85.442 \\ 
    \hline

    {\shortstack{25}}&0.103 &83.750 &92.428 &7.571 &85.048 \\ 
    \hline

        \end{tabular}}
    \label{table:rq3-acc}
\end{table}

\begin{table}[]

    \centering
    \caption{
The table presents a statistical breakdown of the sample counts within each confidence interval, with a total of 3,000 samples in each category. 
    }
     {
  
    \begin{tabular}{c|cccc}
    \hline
     {\textbf{Sample Counts(3000)}} &\multicolumn{4}{c}{\textbf{Confidence Score Intervals}}   \\
     \hline
     
     \multirow{2}{*}{\textbf{Unseen}}& \textbf{(0,0.1)}  & \textbf{(0.1,0.2)} &\textbf{(0.2,0.3)} &\textbf{(0.3,1.0)}  \\
    \cline{2-5}
  
     &2499 &281 &85 &135 
    \\
    \hline
    \hline
     \multirow{2}{*}{\textbf{Seen}}& \textbf{(0,0.7)} & \textbf{(0.7,0.8)} &\textbf{(0.8,0.9)} &\textbf{(0.9,1.0)}\\
    \cline{2-5}
     &213 &81 &106 &2600
    \\
    \hline

        \end{tabular}}
    \label{table:rq3-sample_num}
\end{table}

\begin{framed}
\noindent \textbf{Key Findings: In controlled experiments using \tool to differentiate between seen, unseen, and mixed samples in a target dataset, the AUC achieved was 0.914. Optimal performance was observed with a 20\% FPR, where accuracy and F1 score peaked at 84.750\% and 85.442\%, respectively. For purely seen or unseen datasets, 83.33\% and 86.67\% of samples, respectively, fell within the optimal confidence interval. Overall, these results highlight \tool's robust and effective performance in differentiating samples based on their exposure to the LLM.
}
\end{framed}

\begin{figure}[H]
\centering
\includegraphics[width=0.98\textwidth]{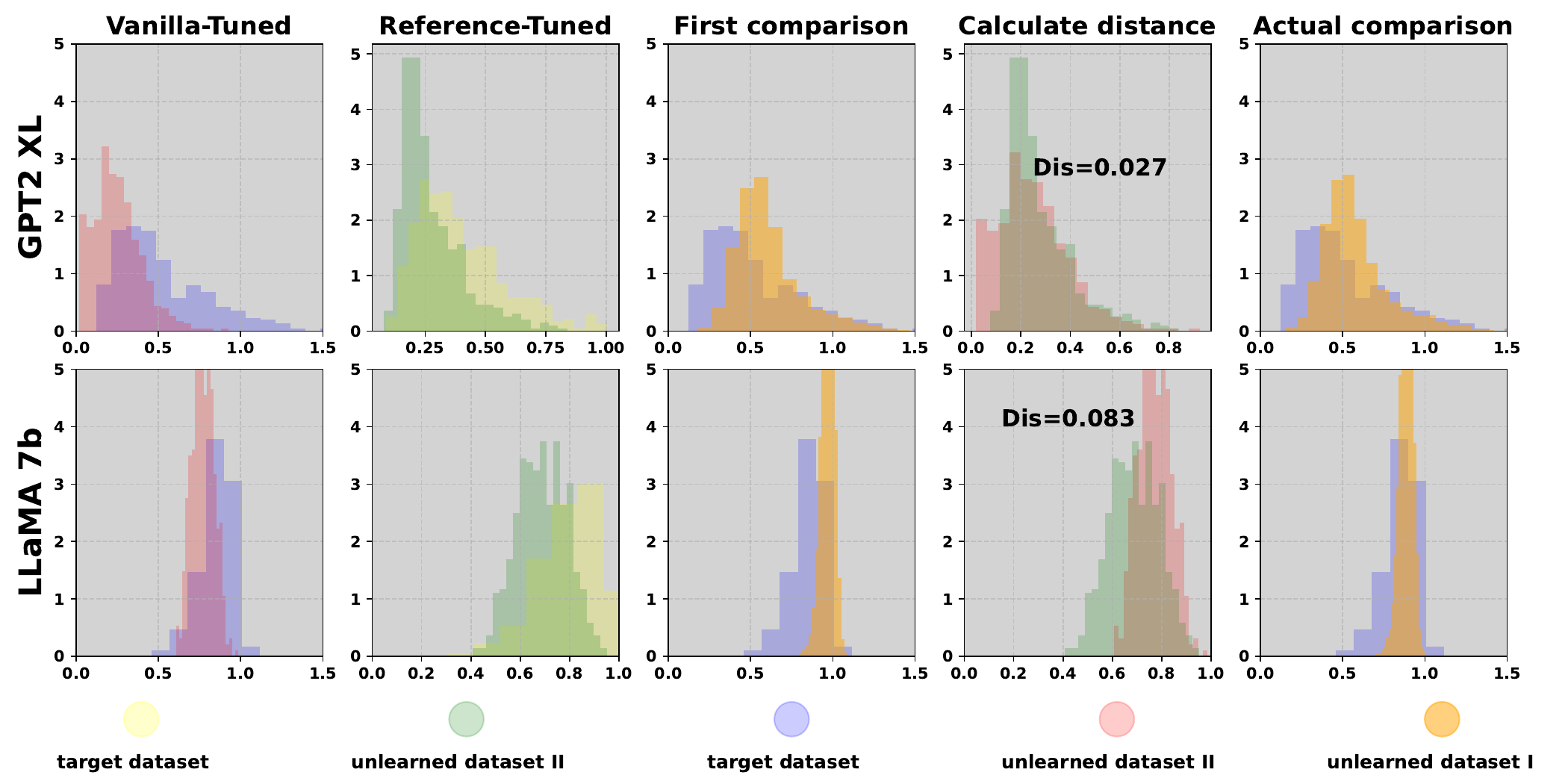}
\caption{In a real-world setting, distribution curves depicting the different distributions of GPT-2 XL (upward) and LLaMA 7b (downward).}
\label{fig:real_world}
\end{figure}

\begin{table}[]
    \centering
    \caption{Classification effects (AUC) for two samples with different GPT-2 versions, different token lengths, and different number of repetitions of the samples. The best results are shown in bold.
    }
     {
  
    \begin{tabular}{cc|cccccc}
    \hline
     \multicolumn{2}{c|}{\textbf{AUC}}  &\multicolumn{6}{c}{\textbf{Number of Token}}  \\
     \hline{}

    \multirow{2}{*}{\shortstack{\textbf{Version}}} &\textbf{repeat}&\multirow{2}{*}{\textbf{50}} & \multirow{2}{*}{\textbf{60}}  & \multirow{2}{*}{\textbf{70}} & \multirow{2}{*}{\textbf{80}} & \multirow{2}{*}{\textbf{90}}& \multirow{2}{*}{\textbf{100}}\\
     &\textbf{times}\\
    \hline
    \hline
   
    \multirow{3}{*}{\shortstack{GPT-2}} 
    &1  &0.67318  &0.70111 &0.72455 &0.74608 &0.76583 &0.78235\\
    &2  &0.76828  &0.80316 &0.83085 &0.85447 &0.87472 &0.89077\\
    &3  &0.84160  &0.87639 &0.90219 &0.92249 &0.93864 &0.95047\\
    
    \hline
    
    \multirow{3}{*}{\shortstack{Medium}} 
    &1  &0.75657  &0.79122 &0.81788 &0.84062 &0.85942 &0.87429\\
    &2  &0.89324  &0.92352 &0.94312 &0.95730 &0.96767 &0.97433\\
    &3  &0.96460  &0.97928 &0.98708 &0.99165 &0.99442 &0.99619\\

    \hline
   
    \multirow{3}{*}{\shortstack{Large}} 
    &1  &0.86596  &0.89626 &0.91749 &0.93277 &0.94408 &0.95222\\
    &2  &0.98733  &0.99291 &0.99532 &0.99673 &0.99748 &0.99804\\
    &3  &0.99919  &0.99952 &0.99964 &0.99969 &0.99974 &0.99975\\
    
    \hline
    
    \multirow{3}{*}{\shortstack{XL}} 
    &1  &0.89705  &0.92303 &0.93964 &0.95218 &0.96107 &0.96670 \\
    &2  &0.99718  &0.99845 &0.99893 &0.99908 &0.99928 &0.99940 \\
    &3  &0.99989  &0.99989 &0.99990 &0.99990 &0.99991 &0.99995 \\

    \hline
   
        \end{tabular}}
    \label{tableRQ4}
\end{table}
\label{RQ1}
To further assess the applicability of our experimental findings, we turned our attention to practical, real-world scenarios. Our objective was to ascertain the existence of samples from a specifically curated target dataset within the training sets of two widely-recognized LLMs: GPT2-XL and LLaMA-7b. To ensure the authenticity of our target dataset and to encompass a diverse range of literary styles, we collated 500 segments of poetry sourced from the internet along with excerpts from globally acclaimed literary masterpieces. The probability density functions derived from these datasets are graphically presented in Figure ~\ref{fig:real_world}.

The subsequent evaluation was divided based on the LLMs in question. Results pertinent to the GPT-2 XL model are depicted in the rows labeled `GPT2 XL', whereas outcomes associated with the LLaMA-7b model can be found in the rows designated as `LLaMA 7b'. To determine the likelihood that each sample in our target dataset was a component of the training data for these LLMs, we computed confidence scores. These scores were derived from the distribution patterns exhibited in the `Actual comparison' column. To gain insights into the frequency distribution of these confidence scores, we undertook a meticulous statistical analysis, categorizing the number of samples that fell into distinct confidence intervals. The intricate details and outcomes of this investigation can be found in Table ~\ref{table:rq4-sample_num}.

\begin{table}[]

    \centering
    \caption{
The table presents a statistical breakdown of the sample counts within each confidence interval, with a total of 500 samples in each category. 
    }
     {
  
    \begin{tabular}{c|cccccc}
    \hline
     \multirow{2}{*}{\textbf{Sample Counts(500)}} 
     &\multicolumn{6}{c}{\textbf{Confidence Score Intervals}}   \\ \cline{2-7}
     & \textbf{(0,0.5)} & \textbf{(0.5,0.6)} &\textbf{(0.6,0.7)} &\textbf{(0.7,0.8)} &\textbf{(0.8,0.9)} &\textbf{(0.9,1.0)}\\
     \hline
     
    {\textbf{GPT2 XL}}&337 &16 &18 &21 &32 &76   \\
    
    \hline
    \hline
     {\textbf{LLaMA 7b}}&299 &21 &19 &42 &35 &84\\

    \hline
        \end{tabular}}
    \label{table:rq4-sample_num}
\end{table}

\begin{table}[]

    \centering
    \caption{
positive samples number 
    }
     {
    \begin{tabular}{c|ccccc}
    \hline
     \multirow{2}{*}{\textbf{Samples Number }} 
     &\multicolumn{5}{c}{\textbf{Threholds(FPR)}}   \\ \cline{2-6}
     & \textbf{0.460(5\%)} & \textbf{0.255(10\%)} &\textbf{0.176(15\%)} &\textbf{0.128(20\%)} &\textbf{0.103(25\%)} \\
     \hline
    {\textbf{GPT2 XL}}&174 &247 &303 &340 &363    \\
    \hline
    \hline
     {\textbf{LLaMA 7b}}&211 &261 &279 &302 &306\\
    \hline
        \end{tabular}}
    \label{table:rq4-thr_pos_sample_num}
\end{table}

\begin{table}[]
    \centering
    \caption{For samples that have been learned by the GPT2-XL a different number of times, a threshold is obtained controlling for the false positive rate of unlearned samples as well as the true positive rate of learned samples selected using this threshold.
    }
     {
  
    \begin{tabular}{cc|cccccc}
    \hline
     \multicolumn{2}{c|}{\textbf{GPT2-XL}}  &\multicolumn{6}{c}{\textbf{Repeat Times}}  \\
     \hline{}

    {\textbf{FPR of Unseen}} & {\textbf{Number of}}&\multicolumn{2}{c}{\textbf{1}} &\multicolumn{2}{c}{\textbf{2}}  &\multicolumn{2}{c}{\textbf{3}} \\
     \textbf{Samples(\%)}&{\textbf{Token}}&{\textbf{THR}}&{\textbf{TPR(\%)}}&{\textbf{THR}}&{\textbf{TPR(\%)}}&{\textbf{THR}}&{\textbf{TPR(\%)}}\\
    \hline
    \hline
   
    \multirow{3}{*}{\shortstack{0}} 
    &80  &1.1409  &2.44 &0.8644 &41.20 &0.5779 &99.04\\
    &90  &1.1333  &2.57 &0.8685 &38.81 &0.5708 &99.08\\
    &100 &1.1251  &2.61 &0.8580 &39.50 &0.5602 &99.27\\
    
    \hline
    
    \multirow{3}{*}{\shortstack{5}} 
    &80  &0.7201  &19.45 &0.5305 &95.18 &0.3352 &99.88\\
    &90  &0.7123  &19.31 &0.5235 &95.51 &0.3332 &99.88\\
    &100 &0.7074  &19.02 &0.5263 &95.19 &0.3316 &99.88\\

    \hline
   
    \multirow{3}{*}{\shortstack{10}} 
    &80  &0.6700  &36.20 &0.4361 &98.27 &0.2743 &99.91\\
    &90  &0.6045  &36.14 &0.4328 &98.33 &0.2736 &99.91\\
    &100 &0.5996  &35.69 &0.4318 &98.32 &0.2696 &99.92\\
    \hline

    \multirow{3}{*}{\shortstack{15}} 
    &80  &0.5401  &55.09 &0.3879 &99.10 &0.2400 &99.92\\
    &90  &0.5351  &54.99 &0.3819 &99.16 &0.2381 &99.93\\
    &100 &0.5292  &55.17 &0.3794 &99.20 &0.2344 &99.94\\
    \hline

    \multirow{3}{*}{\shortstack{20}} 
    &80  &0.4953  &67.80 &0.3522 &99.55 &0.2147 &99.96\\
    &90  &0.4902  &68.08 &0.3469 &99.55 &0.2125 &99.96\\
    &100 &0.4853  &68.38 &0.3445 &99.57 &0.2123 &99.96\\
    \hline

    \multirow{3}{*}{\shortstack{25}} 
    &80  &0.4663  &75.83 &0.3264 &99.68 &0.1956 &99.98\\
    &90  &0.4611  &76.05 &0.3222 &99.72 &0.1961 &99.98\\
    &100 &0.4553  &77.25 &0.3205 &99.77 &0.1961 &99.98\\
    \hline
   
        \end{tabular}}
    \label{tableRQ2-gpt2-xl}
\end{table}

\begin{table}[]
    \centering
    \caption{For samples that have been learned by the LLaMA-7b a different number of times, a threshold is obtained controlling for the false positive rate of unlearned samples as well as the true positive rate of learned samples selected using this threshold.
    }
     {
  
    \begin{tabular}{cc|cccccc}
    \hline
     \multicolumn{2}{c|}{\textbf{LLaMA-7b}}  &\multicolumn{6}{c}{\textbf{Repeat Times}}  \\
     \hline{}

    {\textbf{FPR of Unseen}} & {\textbf{Number of}}&\multicolumn{2}{c}{\textbf{1}} &\multicolumn{2}{c}{\textbf{2}}  &\multicolumn{2}{c}{\textbf{3}} \\
     \textbf{Samples(\%)}&{\textbf{Token}}&{\textbf{THR}}&{\textbf{TPR(\%)}}&{\textbf{THR}}&{\textbf{TPR(\%)}}&{\textbf{THR}}&{\textbf{TPR(\%)}}\\
    \hline
    \hline
   
    \multirow{3}{*}{\shortstack{0}} 
    &80  &1.0994  &0.03 &1.0422 &4.04 &1.0518 &2.79\\
    &90  &1.0941  &0.00 &1.0356 &5.27 &1.0600 &1.23\\
    &100 &1.0827  &0.05 &1.0374 &4.13 &1.0513 &1.92\\
    
    \hline
    
    \multirow{3}{*}{\shortstack{5}} 
    &80  &0.9989  &24.19 &0.9964 &44.91 &0.9937 &58.14\\
    &90  &0.9966  &24.88 &0.9952 &45.47 &0.9926 &59.76\\
    &100 &0.9956  &25.51 &0.9931 &48.38 &0.9911 &62.35\\

    \hline
   
    \multirow{3}{*}{\shortstack{10}} 
    &80  &0.9869  &36.47 &0.9879 &56.55 &0.9856 &69.81\\
    &90  &0.9854  &36.91 &0.9864 &58.49 &0.9844 &71.95\\
    &100 &0.9845  &37.31 &0.9851 &60.50 &0.9830 &73.76\\
    \hline

    \multirow{3}{*}{\shortstack{15}} 
    &80  &0.9792  &44.67 &0.9809 &66.31 &0.9794 &77.45\\
    &90  &0.9777  &45.85 &0.9799 &67.62 &0.9780 &79.47\\
    &100 &0.9767  &46.80 &0.9788 &68.89 &0.9769 &81.11\\
    \hline

    \multirow{3}{*}{\shortstack{20}} 
    &80  &0.9726  &52.12 &0.9756 &72.80 &0.9735 &83.51\\
    &90  &0.9710  &53.93 &0.9745 &73.87 &0.9728 &84.61\\
    &100 &0.9701  &54.73 &0.9736 &74.77 &0.9716 &85.99\\
    \hline

    \multirow{3}{*}{\shortstack{25}} 
    &80  &0.9665  &59.15 &0.9670 &78.18 &0.9687 &87.36\\
    &90  &0.9650  &60.57 &0.9670 &79.49 &0.9676 &88.54\\
    &100 &0.9645  &61.17 &0.9683 &80.30 &0.9668 &89.64\\
    \hline
   
        \end{tabular}}
    \label{tableRQ2-llama-7b}
\end{table}

\begin{framed}
\noindent \textbf{Key Findings: Our evaluation of the 500 poetry segments and literary shows that they are likely to be included within the training of LLMs GPT2-XL and LLaMA-7b. The results highlight the efficacy of our approach in discerning the origin of diverse textual samples in real-world scenarios.}

\end{framed}
\section{Discussion}\label{sec:discussion}

\textbf{Cost for Training and Prediction.} Based on the difference in sample loss before and after fine-tuning the LLMs, we can effectively distinguish between samples that the target LLM has learned and those not. However, in order to obtain the loss differences for these samples and establish the baseline distribution for evaluation, we selected a substantial number of samples that the LLM has not learned and fine-tuned the target LLM multiple times on these samples. Additionally, inference on these samples is also required. Therefore, these methods entail significant computational costs and increased storage resources. Although our study is primarily concentrated on small to medium-sized LLMs, the associated costs do not substantially compromise our ability to obtain meaningful results. Meanwhile, enhancing detection efficiency represents a promising avenue for future research

\noindent\textbf{Target Probability Calculation.} In our study, we calculate the probability that a sample in the target dataset belongs to the set of learned samples by examining the distribution of loss differences in the unlearned dataset I. We operate under the empirical assumption that this distribution adheres to a normal distribution. Concurrently, we are also investigating the utility of kernel density estimation as an alternative approach. A comparative analysis of the effectiveness of these two methodologies will be a focal point in our forthcoming experiments.

\section{Threats To Validity}
\subsection{Internal Threats}
\noindent\textbf{Lack of Ground Truth Labels.} Owing to the undisclosed nature of the training datasets for GPT-2 and LLaMA, direct validation of their contents remains infeasible. To mitigate this challenge, we have employed a validation strategy that utilizes samples sourced from the web, which are likely candidates for inclusion in these training sets. Although our empirical findings align well with our theoretical expectations, the unavailability of verified labels for these datasets introduces a notable limitation to our study. 

\noindent\textbf{Limited LLMs Included.}One internal threat to the validity of our findings is the limited variety of LLMs selected for the study. Specifically, we chose to focus on GPT-2 and Llama7b, which are relatively small to medium-sized LLMs. Consequently, the scope of identified copyright violations might be narrower than what could be observed in larger LLMs like GPT-3, Llama13b, or Llama30b. The rationale for this selection was primarily computational; fine-tuning larger models demands significantly more computational resources, making it a more expensive endeavor both in terms of time and financial investment. Therefore, while our findings provide valuable insights into the nature of copyrighted content in LLMs, they may not fully capture the extent of the issue in larger, more complex models.

\subsection{External Threats}
\noindent\textbf{Limited Confidence Level Calculation.}
Regarding the computation of confidence levels in our study, we employed normal distribution fitting as our method. While this is a commonly used technique, it is not the only approach for calculating confidence. Alternative statistical methods, such as Bayesian inference or bootstrapping, could potentially offer different insights into the reliability of our findings. The choice of normal distribution fitting could, therefore, be seen as a limitation in our methodology, affecting the external validity of our results.

\noindent\textbf{Copyright Legal Consideration.}
Another external threat is the jurisdictional and cultural context in which our study is situated. Copyright laws can vary significantly between countries, and what constitutes a violation in one jurisdiction may not be considered the same in another. Our study is based on a specific legal framework, which may not be universally applicable, thereby limiting the global generalizability of our findings. However, our work is mainly focus on the detection technique for copyright, which is orthogonal to this aspect. 
\section{Related Work}\label{sec:literature}

\subsection{Sensitive Information Analysis for LLMs}
LLMs have been demonstrated through various research to possess memorization capabilities~\cite{petroni2019language}, retaining privacy and sensitive information present in their training datasets. 
When prompted appropriately, LLMs exhibit a tendency to verbatim reproduce learned content, sentence by sentence. This is evidently undesirable, as it not only leads to the potential leakage of privacy and sensitive information but also results in a degradation of the quality of generated content. Carlini et al.~\cite{carlini2022quantifying} has demonstrated that longer contextual prompts, increased exposure to learned samples, and greater model capacity all contribute to LLMs being more prone to verbatim reproduction of previously learned content. 
Previous research~\cite{carlini2021extracting,lukas2023analyzing,lee2023language,vakili2021clinical,rocher2019estimating} has indeed demonstrated that sensitive information such as URLs, personal details, and phone numbers can be reconstructed. Inan et al.~\cite{inan2021training} studied the recovery of individual sentences. 
To the best of our knowledge, the research conducted by Chang et al.~\cite{chang2023speak} is currently the sole study addressing the issue of book copyright in training datasets for Large Language Models (LLMs). They evaluated the performance of ChatGPT and GPT-4 in terms of their familiarity with known books through a cloze task.

\subsection{Member Inference Attacks Against LLMs}
Member inference attacks refer to the assessment of whether a target sample has been part of the training data for the target LLM~\cite{shokri2017membership,yeom2018privacy}.Many previous studies have been based on the loss values of samples on the target model~\cite{shokri2017membership,yeom2018privacy,song2020information}, determining sample membership by employing a threshold approach.
Recent research by Mireshghallah et al. ~\cite{mireshghallah2022quantifying} has shown that using sample loss values on the target LLM as an indicator is incorrect. This is because samples exhibit false variations in loss due to their inherent complexity. Our work is inspired by this notion.

\section{Conclusion}\label{sec:conclusion}

In this paper, we introduce a universal optimization framework, denoted as \tool, designed to test whether a sample has been learned by the target LLM. We conduct a thorough feature study to understand the characteristics of sample loss and the rate of change in sample loss as samples are learned by the LLM. Based on these characteristics, we formulate the difference in loss change as an indicator to distinguish between samples that have been learned by the LLM and those that have not.

To establish a baseline distribution, we perform experiments on vanilla LLM. Subsequently, we fine-tune the reference LLM and vanilla LLM with the test samples, leveraging the differences between these two distributions to refine the baseline distribution, which serves as the criterion for classification.

Extensive experiments demonstrate that \tool effectively identifies samples that have been learned by the target LLM. Our research unveils the promising prospect of utilizing differences in sample loss change to identify samples learned by the LLM, thus paving the way for exploring copyrighted novels that have been learned by LLM, opening up a vast avenue of opportunities in this domain.

\bibliographystyle{ACM-Reference-Format}
\bibliography{bib}

\appendix
\section{APPENDIX }
 \label{appendix:pre_study_book_list}
 
\begin{table}[H]
    \centering
    \caption{Five books published in 2023, selected in Section~\ref{sec:pre_RQ1}, along with their genre information. 
    }
     {
    \begin{tabular}{c|c|c|c}
    \hline
     {\textbf{Name of Novels}}  &{\textbf{Author}} &{\textbf{Main Genres}} &{\textbf{Publication Date}}\\
     \hline
     \hline
    Happy Place  &Emily Henry  &Contemporary & April 25, 2023
 \\ 
    Clytemnestra    &
Costanza Casati &Mythology &May 2, 2023\\ 
    Godkiller(\#1)   &Hannah Kaner &Fantasy  &January 19, 2023\\ 
    Said No One Ever  &Stephanie Eding
&Romance &April 4, 2023\\
    Some Desperate Glory &
Emily Tesh &Science Fiction &April 11, 2023\\
    \hline
    
        \end{tabular}} 
   
\end{table}

\end{document}